# Understanding strong $H_2$ emission in astronomical environments

**Gary J. Ferland[1]**
*Physics, University of Kentucky*
*Lexington, KY, 40506 USA*
*E-mail:* `gjferland@gmail.com`

Here I describe recent studies of objects with $H_2$ emission that is strong relative to other spectral lines. Large telescopes and fast spectrometers have made the 2 μm window accessible even for relatively faint objects. I summarize several environments where strong $H_2$ 2.121 μm emission is observed. The line is hard to excite due to its large excitation potential, and is most emissive in regions that have temperatures that are nearly high enough to dissociate $H_2$. I outline several case studies. In the Helix planetary nebula strong emission is produced by rapidly flowing molecular gas that is exposed to an intense ionizing radiation field. This advective production of $H_2$ is a fundamentally non-equilibrium process. In the filaments surrounding brightest cluster galaxies in cool core clusters ionizing particles penetrate into magnetically confined molecular cores and excite the gas. Finally, I outline ongoing work on the Crab Nebula, where the first complete maps of molecular emission have only recently been completed. Both ionizing particles and high-energy photons may be important. Finally I speculate on the origin of the correlation between $H_2$ / H I intensity ratios and other properties found in Active Galaxies. This is suggestive of a hardening of the radiation field along the Eigenvector 1 sequence. In all of this work I take the approach of understanding $H_2$ emission along with emission from low and moderate ionization species, a necessary step if we are to really understand the context in which $H_2$ emission forms.

---

[1]  Speaker

# 1. Introduction

The opening of the 2 μm spectral window, with large-aperture telescopes and fast spectrometers, has made the study of the $H_2$ and H I emission lines within this spectral range common, even in relatively faint sources. $H_2$ emission is fairly ubiquitous. Indeed, strong $H_2$ emission is so commonly encountered that it is often taken for granted without asking what conditions or energy sources might cause it. While this paper will focus on the different mechanisms that make strong $H_2$ emission, I especially want to understand the correlation found in [1] between $H_2$ emission and other spectra properties of starburst galaxies, AGN, and LINERs. A cartoon of one of their figures, showing that $H_2$ becomes progressively stronger along the sequence, is shown in Figure 1. Some of the objects discussed below have been placed on this figure. What does this tells us how activity near the black hole communicates with molecular gas?

This summary is based on a series of papers written over the past half dozen years. Many of the results were obtained with the plasma simulation code Cloudy [2] that was extended to include molecules as parts of the PhD theses of Gargi Shaw [3] and Nick Abel [4]. [5] summarizes a workshop which brought together developers of many different codes to compare predictions for "PDRs", the $H^0 - H_2$ transition layer that lies between an H II region and its associated molecular cloud. The big difference between the approach we take with Cloudy, versus a conventional PDR code, is that we simultaneously predict emission from molecules and *all* ions of the lightest thirty elements. In particular we predict the full H I emission spectrum, a tracer of $H^+$. It is relatively simple to predict $H_2$ by itself, and obtain good fits with observations, but it is far more challenging to try to understand the full spectrum. We take the latter approach because this tells us far more about the astronomical context, and what is happening in front of our telescopes.

# 2. What $H_2$/H I emission ratio is expected for a star forming region?

I focus on the ratio of the $H_2$ 2.121 μm line relative to nearby H I recombination lines, either Brγ or Pα. To understand what their ratio means we must first consider how each line forms.

A typical ionization structure, where starlight strikes a molecular cloud, is shown in Fig 8.4 of [6] (hereafter AGN3). A series of layers in which H occurs as $H^+$, $H^0$, then $H_2$, surrounds the stars. The H I lines form by recombination in the ionized region. Their intensity is proportional to the number of hydrogen-ionizing photons emitted by the stars or AGN, as described in Chapter 2 of AGN3. Note that a typical spectral energy distribution (SED) of a young star cluster or an AGN peaks at hydrogen-ionizing energies, so strong H I lines are produced.

Figure 2 shows this graphically. It gives the SED of typical AGN and is adopted from Figure 13.7 of AGN3. The long rightward-pointing arrow shows the range of photons that are capable of ionizing hydrogen. The energy in this part of the SED is converted into various emission lines, including H I, when light strikes matter. The resulting lines are very strong.

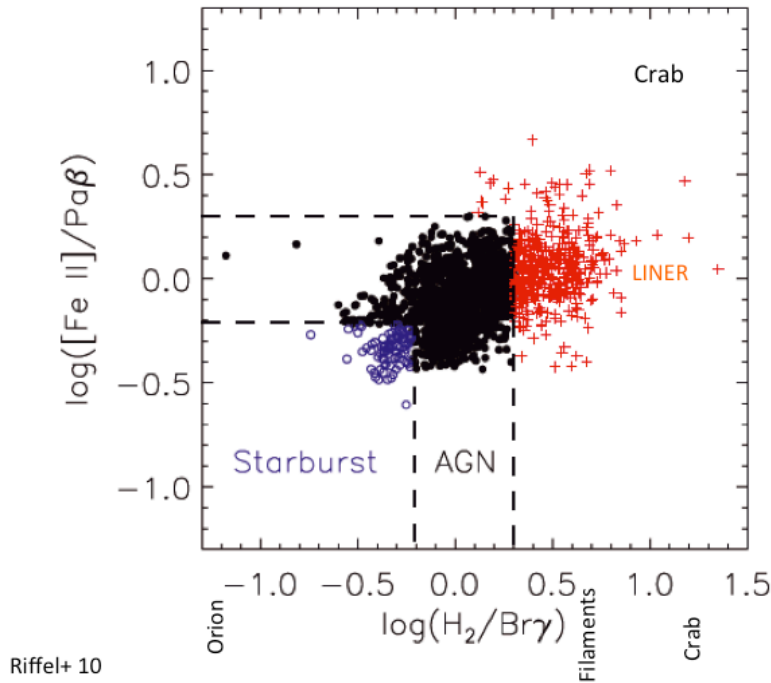

Figure 1 This shows the $H_2$ / H I ratios in a number of objects, along with [Fe II] / H I. The ratios observed in the Orion star-forming region, a filament in a cool core galaxy cluster, and the Crab Nebulae, are also shown. Adopted from Riffel et al. (2010, [1]).

$H_2$ lines are far more difficult to form because of the wide energy separation of the energy levels. Figure 3 (Figure 8.3 of AGN3) shows the lowest rotation and vibration energy levels. The 2.121 μm transition is shown and the upper level has an energy of ~7000 K above ground. This has the important implication that the line can only be collisionally excited in gas that has kinetic temperatures approaching the dissociation energy of the molecule.

In a typical H II region and PDR, hydrogen becomes molecular only in cold regions that are well shielded from dissociating radiation. The gas kinetic temperature in these regions is low, typically less than 100 K, because there is little light to heat the gas. This means that direct collisional excitation is not an important $H_2$ line formation process. The $H_2$ lines are formed by the continuum fluorescence process discussed, for instance, in AGN3 or [7]. In this process a fraction of the stellar continuum around 1000Å is converted into $H_2$ line emission. The small part of the SED that can pump $H_2$ lines is shown in Figure 2. The resulting $H_2$ lines are weak compared to H I recombination lines because, for an SED similar to an O star or AGN, there are relatively few photons in the ~1000Å band that produces $H_2$ emission, compared with the number of hydrogen-ionizing photons which produce H I emission. The result is the small ratio of $H_2$ / H I indicated for the Orion star-forming region in Figure 1.

The continuum fluorescence process does not depend on the local gas kinetic temperature, so the $H_2$ level populations that result are fairly independent of it. As a result the $H_2$ excitation diagram, in which the level population is plotted as a function of excitation energy, does not indicate a single temperature, and the range of apparent temperatures does not reflect the kinetic temperature of the gas. This is discussed, for instance, in [8].

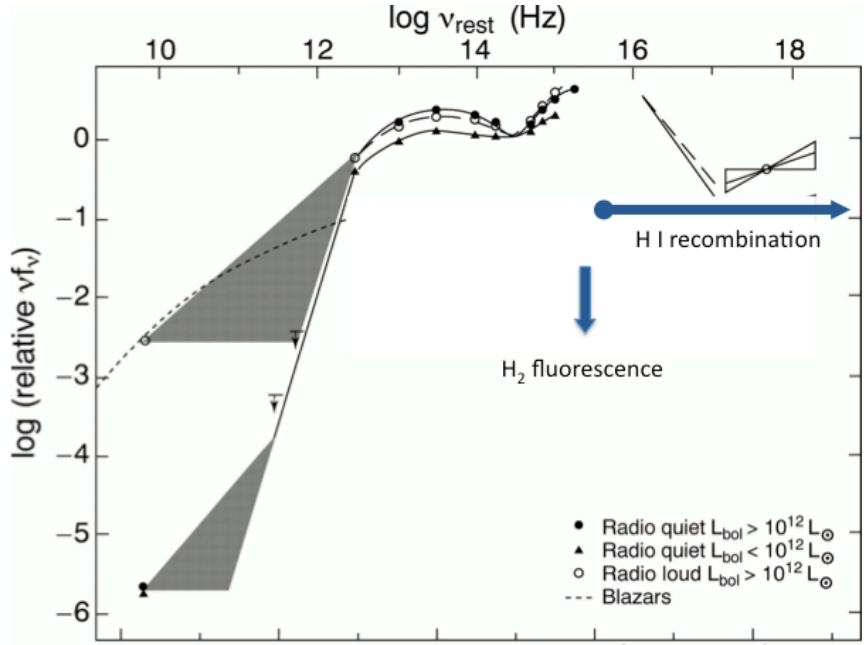

Figure 2 This shows the SED of typical active galactic nuclei. The long right-pointing arrow shows the range of the radiation field that is converted into H I emission. The width of the downward pointing arrow shows the range of the SED that can produce H₂ fluorescent emission. Adopted from AGN3.

To summarize, in this H II region – PDR limit, in which light striking the cloud produces layers of $H^+$, $H^0$, and $H_2$, we expect an $H_2$ / H I ratio that depends mainly on the form of the SED. Small ratios are expected when the bulk of the photons occur at ionizing energies. Large ratios are produced, for instance, in reflection nebulae, where the stars are too cool to produce much hydrogen-ionizing ration. The resulting $H_2$ level populations are produced by a fluorescent excitation process that creates a non-thermal distribution.

## 3. Why hot H₂ is common

Figure 1 shows that very large $H_2$ / H I ratios are observed. The figure also shows some ratios seen in filaments surrounding the brightest galaxy in cool-core clusters [9] and the Crab Nebula [10]. The $H_2$ / H I ratio is roughly 2 dex larger than seen in a typical H II region / PDR. $H_2$ level temperatures can also be measured in these cases, resulting in surprisingly high values [11, 12]. Similarly warm $H_2$ is seen in shocks around HH objects.

These high $H_2$ temperatures are likely to be real, and reflect the gas kinetic temperature. Indirect $H_2$ excitation processes such as grain formation pumping, or the continuum pumping outlined above, are not capable of producing $H_2$ emission as strong as is observed. Collisional excitation at warm temperatures is the easiest way to make such strong $H_2$ emission. $H_2$ emission can be arbitrarily strong if the molecular gas can be heated to the high temperatures indicated by the level populations.

These high temperatures are almost certainly the result of the selection effects identified by [13] and [10]. Figure 4, taken from [10], shows the emissivity of the $H_2$ 2.121 μm line as a function of gas density and kinetic temperature. In these calculations the gas is well shielded from light that could pump or dissociate the molecule. That the molecular equilibrium is set by the balance between grain catalysis formation (ISM formation rates were assumed) and destruction by cosmic rays (the galactic background is assumed) and by thermal collisions with gas constituents.

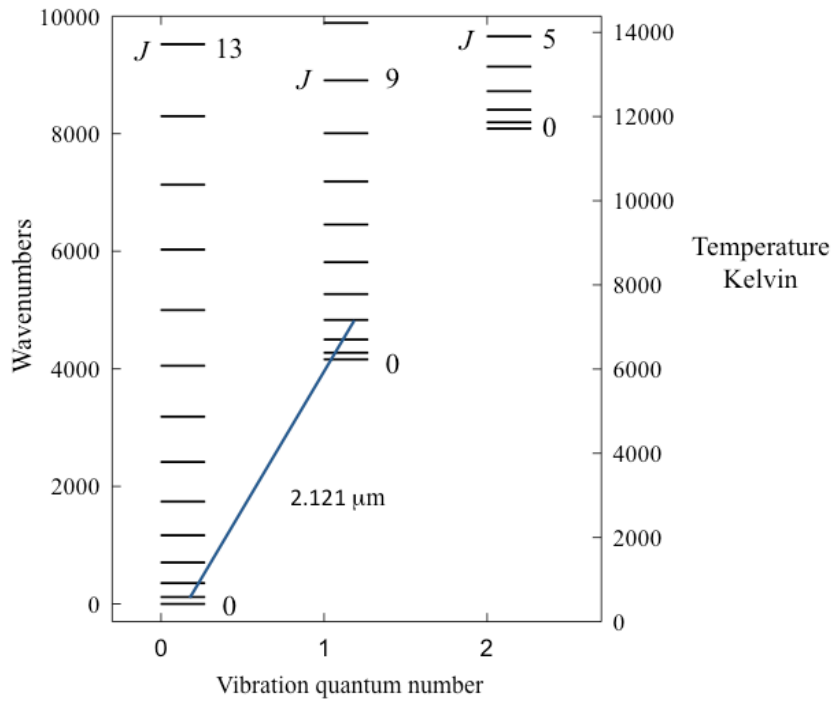

Figure 3 – the lower rotation and vibration energy levels of $H_2$. The diagonal line shows the 2.121 μm transition that is the subject of this paper. The upper level has an excitation potential of roughly 7000 K, making it very difficult to excite except in very warm gas. Adopted from AGN3 and [3].

The contours show that the emissivity is strongly peaked at ~3000 K, the temperature deduced in the strong $H_2$ objects. This is almost certainly due to the selection effects introduced by the physical processes that determine how $H_2$ forms and emits. Most hydrogen is molecular at low temperatures, because collisional destruction processes are slow, but the gas is also too cool to excite much 2.121 μm emission. As the temperature increases the Boltzmann factor that determines the excitation becomes larger, and the line emissivity increases. For temperatures above the peak the emissivity goes down as $H_2$ is dissociated and the gas becomes atomic and eventually ionized.

In a real interstellar cloud, gas could be present with a wide range of densities and temperatures. We will detect gas with the mix of density and temperature leading to maximum emissivity. This is the Locally Optimally-emitting Cloud ("LOC") model that has been applied successfully to AGN emission line regions [14]. The central question that remains then is what produces such high temperatures?

## 4. How to use plasma simulations to understand the spectrum

Here I outline my philosophy for understanding the spectra of astronomical sources, often with the aid of spectral simulations such as those done by Cloudy. We are trying to solve the inverse problem – we know the answer – the observed spectrum – and we are trying to understand the question – what conditions caused the gas to produce this spectrum?  It is often not possible to go back unambiguously to the correct conditions, just as the number 42 is the solution of many different equations.   It is necessary to take an approach that is physically motivated and most likely to lead to progress [15].

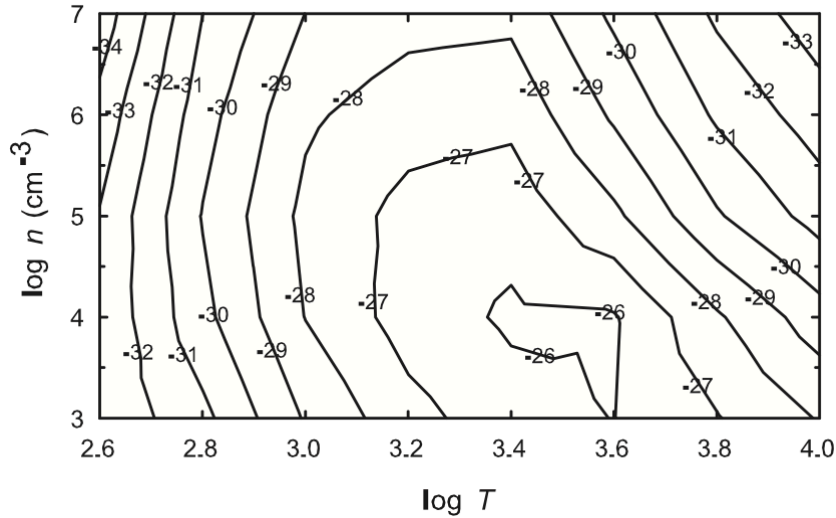

Figure 4  The log of the volume emissivity for the H$_2$ 2.121 μm line, adopted from [10].  The logs of the kinetic temperature and hydrogen density are the two axes.  The emissivity is very strongly peaked to temperatures around 3000 K due to the high excitation potential of the lines, as shown in Figure 3.

### 4.1 Free parameters – the curse of any model

Most physical models for an astronomical process will have free parameters.  This is inevitable since we can seldom trace a physical process all the way back to fundamental first causes.  For instance, the gas density and the flux of ionizing photons striking the gas determine the conditions in a photoionized cloud.  The velocity sets the conditions in a shock.  These are a tractable problem because there are many observables (the emission-line spectrum) that result from these parameters.

I have seen papers where the number of free parameters greatly exceeds the number of observations.  HH objects often have an H$_2$ excitation diagram which could be described as a second order polynomial. A range of shocks may be used to fit these lines. Each shock component brings with it several free parameters, and the number of free parameters can easily outnumber the three-parameter polynomial that describes the observations..

For a model to have meaning there must be fewer free parameters than observables.  Such a model does have predictive powers and does test a physical picture. But using a model with $N$ parameters to fit $n<N$ observations proves nothing – it is a mathematical tautology – the model must work and its success says nothing about what is happening in front of our telescopes.

In many cases boundary conditions obtained from other knowledge of the astronomical environment can be used to remove free parameters.  There are even environments, such as filaments in the Helix or Crab Nebulae described below, where so much is known that there are essentially no free parameters.

**4.2 Concerning conspiracies and different energy sources**

It seems unlikely, to me, that independent energy sources will conspire so that each makes a significant contributions to single spectrum. For instance, in the case of the $H_2$ / H I ratio that is the focus of this paper, it might be argued that photoionization produces the H I emission lines while shocks produce the $H_2$ lines. It would require a remarkable coincidence for two independent processes to inject about the same amount of energy. A spectrometer is a linear detector – for it to notice ionized and molecular components in the same spectrum these regions must emit roughly equally, or perhaps in a 10:1 proportion.

Table 1 compares mechanical and radiative heat input in two very different places – the Earth and a typical late phase supernova remnant. For the Earth there is a bow shock caused by our orbit through the solar wind although sunlight is the dominant heating process. For the SNR galactic background starlight photoionization does occur in addition to shock heating. In both regions a single energy source dominants by a large fraction. I know of no well-observed environment where independent energy sources contribute equally to heating the region.

| Table 1 Heating rates [erg cm$^{-2}$ s$^{-1}$] in two astronomical environments | | |
|:---:|:---:|:---:|
| Object | Radiative heating | Collisional heating |
| Earth | 1 400 000 | 5.8 |
| SNR | 0.0028 | 7.9 |

What adding ad hoc energy sources does do is increase the number of free parameters. It follows trivially that a closer fit to the spectrum will result, when measured in terms of RMS deviations between the final model and the observations. But a more useful metric is $\chi^2$, which takes into account the number of free parameters.

It may be possible to devise a scenario where some underlying physics would couple different processes – for instance, the ISM is observed to have rough energy equipartition between several different energy sources, and in the case of the Crab Nebula, the ionizing synchrotron radiation field is produced by ionizing particles (the relativistic electrons) so the two should be coupled. But it seems more physical, to me, to begin with the assumption that a single energy source dominates.

**5. Case studies – how to get strong $H_2$ emission**

This section describes a number of astronomical environments where strong $H_2$ emission is seen. For this to happen there must be hot molecular gas, heated to nearly its dissociation temperature, and this must correlate with an $H^+$ region. In this picture there could be large reservoirs of cold and undetectable gas. There may be transient regions, where gas is heated to its dissociation temperature and emits while it passes through the "magic" temperature. Once heated enough, $H_2$ is destroyed, becoming $H^0$ and eventually $H^+$.

This is only meant as short description of the physics operating in a few well-studied regimes. The original literature gives more details.

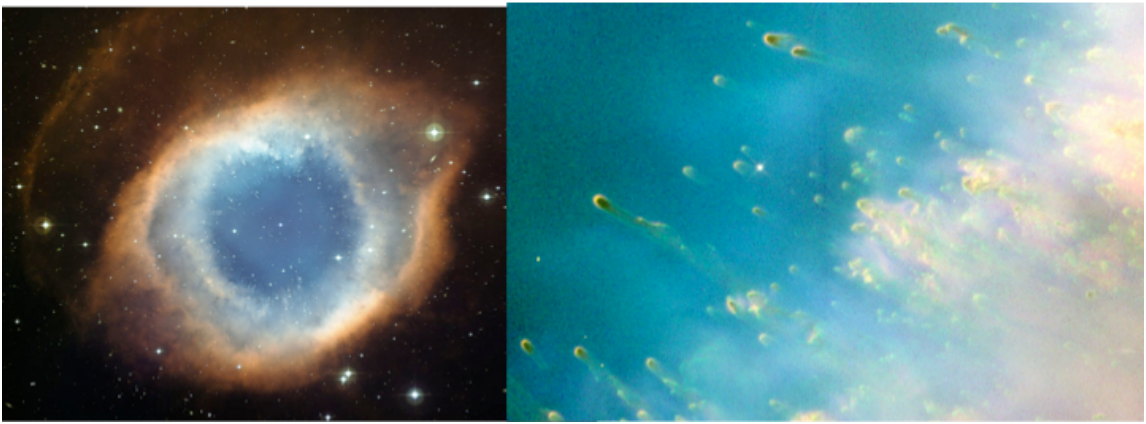

Figure 5 Two HST images of the Helix planetary nebula. The image on the left shows the main nebula while the zoom on the right shows the molecular clumps described in the text. Adopted from the Hubble Heritage site.

### 5.1 The Helix planetary nebula

Figure 5 shows the famous HST image of the Helix, a young and well-studied planetary nebula (see Chapter 10 of AGN3 for an overview). Most of the envelope was ejected from an AGB star within the last $10^4$ years. It is thought that dense clumps, shown on the right, formed during this dense ejection phase. These clumps have strong molecular emission, including prominent $H_2$ lines.

The Helix is an excellent case study because many of its properties are known from other observations. For instance, the composition of the gas, the temperature and luminosity of the central star, and the projected separation between the clumps and the central star, are all reasonable well understood. The conditions and spectrum of the nebula are the result of photoionization by the radiation field emitted by the central star.

Photoionization simulations in which the gas is assumed to be static and in equilibrium cannot come close to producing the strong $H_2$ lines, relative to H I lines, seen in the clumps. The SED from the central star has even more H-ionizing photons, relative to the 1000Å continuum, than an AGN, so the $H_2$ / H I ratio is expected to be smaller than one would estimate from Figure 2. Simulations do predict that the cores of the clumps seen in the right panel of Figure 5 should be molecular, but the gas is expected to be too cool to produce much 2.121 μm emission.

The images suggest that the clumps are ablating material into the surrounding ionized gas. [16] showed that the outward flow of material allows some molecular hydrogen to be present in regions which, in the static case, would be ionized. In this case molecular gas flows outward from the clump core. The gas density decreases as the gas accelerates and expands outward. Molecular gas quickly moves into regions where ionizing radiation is present. The gas is heated and $H_2$ survives for a long enough time to produce strong emission before dissociating and eventually becoming ionized. During its voyage from the cold molecular cores into hot ionized regions its temperature passes through the magic temperature, shown in Figure 4, where strong $H_2$ emission is produced.

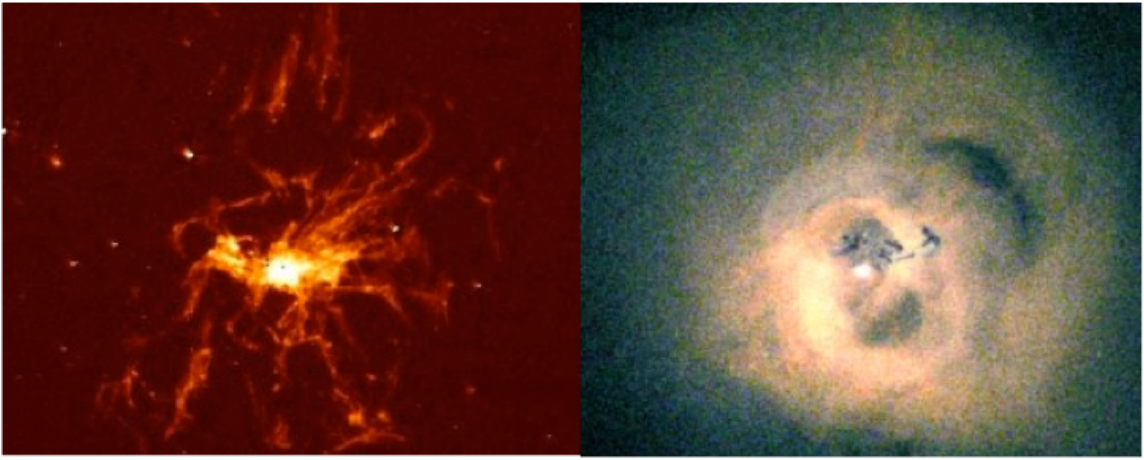

Figure 6 The optical filaments around NGC 1275 in the Perseus cluster are shown on the left and an X-ray image on the same scale is on the right. Adopted from [17] and [18].

In these blobs the strong H$_2$ emission is a consequence of gas, irradiated by a hot planetary nebula central star, boiling off from molecular clumps embedded in a lower density hot ionized gas. In effect, both the H$_2$ and H I emission are produced by the same ionizing radiation field.

**5.2 Filaments in cool-core clusters of galaxies**

Filaments are often observed to surround the brightest cluster galaxy (BCG) in cool-core clusters of galaxies. Figure 6 shows an Hα image of the Perseus cluster on the left. The filaments have remarkable optical emission-line spectra, with strong [N I] λ5198, and other low-ionization species. Strong H$_2$ emission is seen in the IR and large masses are deduced from CO radio observations. The observational literature is vast and is summarized in [11], [9], and [17].

Figure 6 shows an X-ray image on the right. The filaments are immersed in a hot high-pressure plasma. Star formation is clearly occurring in regions close to the BCG but the distant filaments, which have been the focus of our work, has no strong evidence for a stellar component. Other sources of light such as the X-ray emission from the intracluster medium or the central AGN cannot account for the brightness of the filaments or their odd spectrum. What energizes them?

[9] concluded that ionizing particles were responsible for heating, exciting, and dissociating/ionizing the filaments, and [18] showed that the hot intracluster medium shown in the right panel of Figure 6 produces the required flux of particles. When energetic particles penetrate into an atomic or molecular gas they produce a shower of secondary particles that cause heating, excitation, and dissociation. In such an environment ionization is predominantly produced by the secondary electrons and thermal collisional excitation, while recombination is driven by hydrogen charge exchange due to the large abundance of H$^0$. These peculiarities account for the odd optical emission-line spectrum.

The picture that emerges is one where predominantly molecular gas is present in large reservoirs. The filaments are confined by the magnetic field that accounts for the geometry. Gas is free to move along field lines and maintain constant gas pressure. The result is that dense regions are cold and molecular while low-density regions are warm and ionized. All have

page 10

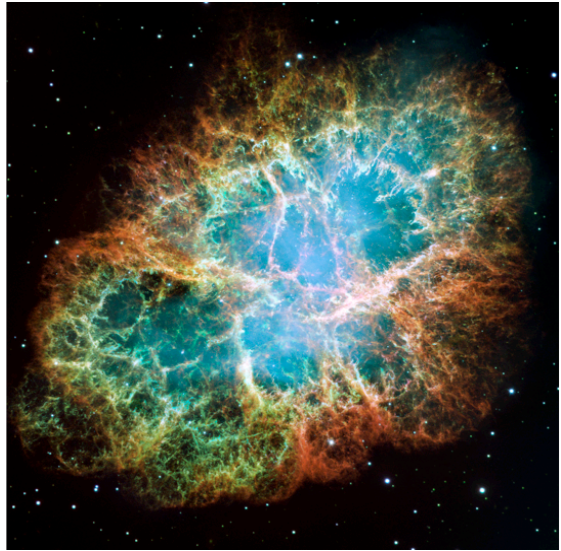

Figure 7 The HST PR image of the Crab Nebula.  $H_2$ emission correlates with low-ionization emission and is located along the prominent filaments.  Adopted from Hubble Heritage.

conditions set by the density – temperature pair required to maintain constant pressure, and a flux of ionizing particles needed to account for these conditions.  The theory has a single free parameter, a power-law index accounting for the gas density distribution.  The intensities of a large number of ionic, atomic, and molecular emission lines were reproduced [9].

**5.3 The Crab Nebula**

The familiar HST image of the Crab is shown in Figure 7 (Chapter 12 of AGN3 gives an overview of such remnants).  We began working on the Crab as a nearby laboratory to use as a proxy for the far fainter filaments in cool-core galaxy clusters.  It is a very young SNR that is not yet interacting with the ISM.  The radiation field emitted by the synchrotron gas should dominate conditions in the filaments.  Ionizing particles are also present since they produce the synchrotron emission.  A large number of studies have concluded that the emission from the filaments is produced by photoionization by the synchrotron continuum.

[19] discovered strong $H_2$ emission in a part of the Crab.  I began a large study of the molecular content of the filaments in collaboration with astronomers at Michigan State University using their SOAR 4m telescope.   A first paper [10] discovered very strong $H_2$ emission in a region close to the position discussed in [19].  We argued that a significant amount of mass must be associated with the molecular clumps.  A survey covering much of the Crab showed that strong $H_2$ emission is pervasive and tends to be associated with strong low-ionization optical emission, especially the [S II] doublet near H$\alpha$ [20].  A number of additional $H_2$ lines were detected in [12].  Excitation temperatures around 3000 K were found, in agreement with the LOC prediction shown in Figure 4.

The [S II] lines that correlate with $H_2$ are produced by photoionization by the synchrotron continuum.  The Crab is a remarkably hostile environment – both this hard continuum and ionizing particles are present.  This is still a work in progress with no model now in hand that accounts for the full observed ionic and molecular emission.  The Crab may be an example where two seemingly different energy sources, photoionization and ionizing particles, are coupled by underlying physics. The synchrotron continuum that produces the photoionization is itself produced by relativistic electrons moving in a magnetic field. There are many open questions.

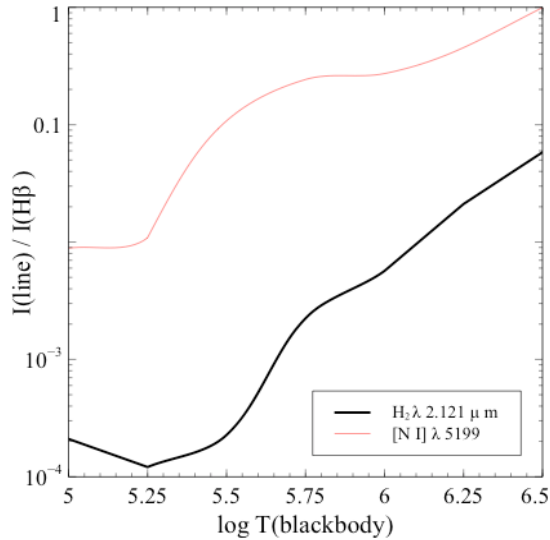

Figure 8 A simple calculation showing the effects of irradiating a dusty cloud with an increasingly hard SED. A blackbody is used for simplicity and only two line ratios are shown, the $H_2$ 2.121 μm and [N I] λ5198Å relative to Hβ. The $H_2$ / H I ratio becomes larger as the SED becomes harder since more energetic photons are able to penetrate into neutral or molecular gas.

## 6. The Starburst - AGN – LINER sequence – a changing SED?

What does all this say about the Riffel sequence shown in Figure 1? The modeling described above shows that strong $H_2$ emission can be produced when high-energy photons, or particles, penetrate into molecular gas, or when molecular gas rapidly flows into regions where ionizing photons are present. The "fewest energy sources" argument given in Section 4.2 suggests that we should try to account for the sequence with only photons, since the bulk of the optical and UV lines seen in AGN are produced by photoionization.

Many properties of the black hole, including mass and accretion rate, may change along the Starburst / AGN / LINER sequence. The SED emitted by the black hole and accretion disk is likely to change as a result. As a test I did a series of photoionization calculations in which the SED of the ionizing continuum was varied. For simplicity I assumed a blackbody and that the gas has ISM abundances and grains, a hydrogen density of $10^4$ cm$^{-3}$, and an ionization parameter of $U = 10^{-2}$. The black body temperature was varied to show the effects of an increasingly hard SED.

The result is shown in Figure 8, which shows how the $H_2$ and [N I] emission vary relative to H I. As the SED becomes harder there are relatively more high-energy photons that are able to penetrate beyond the $H^+$ region, where H I lines form, and enter atomic and molecular gas. These produce strong $H_2$ emission as the molecular gas absorbs XUV and X-ray photons. Such a scenario is capable of producing correlations like those shown in Figure 1.

## 7. Conclusion

This has been a fairly wide-ranging survey of recent work on $H_2$ emission in several astrophysical environments. The central challenge is to get enough energy into molecular gas to cause it to emit in high excitation potential lines such as the 2.121 μm line. There are several ways to do this – shocks have been covered extensively in other work and are not described here. Penetrating ionizing photons and particles are another way. Finally, a rapid flow that causes molecules to enter regions where they should not, in equilibrium, exist, also produces strong $H_2$ emission.

AGN are likely to be photon dominated, and the previous section showed a simple calculation in which the Riffel correlation in Figure 1 might be accounted for by changes in the hardness of the SED. [Fe II] is also part of this correlation. Optical Fe II emission is the strongest indicator of eigenvector 1 in PCA analysis. Could eigenvector 1 reflect changes the SED? The results are suggestive and an exceptionally hard SED may explain some of the spectroscopic peculiarities the NLSy1 such as strong Fe II, [N I], and $H_2$ emission. It is suggestive that the Crab Nebula, which is also plotted in Figure 1, has both very strong [Fe II] and $H_2$ emission.

I want to thank my $H_2$ collaborators Jack Baldwin, Andy Fabian, Roderick Johnstone, and Ed Loh, and Philippe Salome for many helpful discussions. Support by NSF (0908877), NASA (07-ATFP07-0124, 10-ATP10-0053, and 10-ADAP10-0073) and STScI (HST-AR-12125.01 and HST-GO-12309) are gratefully acknowledged.